\documentclass{iau}
\usepackage{graphicx,natbib}
 
\newcommand{\apj}{ApJ}           
\newcommand{\apjl}{ApJ}           
\newcommand{\mnras}{MNRAS}       

\newcommand{\apjs}{ApJS}           

\title{Probing Individual Star Forming Regions Within Strongly Lensed Galaxies at z $>$ 1}

\author[Bayliss et al.]{Matthew B. Bayliss$^1$, Jane R. Rigby$^2$, Keren Sharon$^3$, Michael D. Gladders$^4$ \and Eva Wuyts$^5$}

\affiliation{$^1$Dept. of Physics, Harvard University, 17 Oxford St., Cambridge, MA, 02138 \\ email: {\tt mbayliss@cfa.harvard.edu} \\
$^2$Obs. Cosmology Lab, NASA Goddard Space Flight Center, Greenbelt, MD, 20771 \\
$^3$Dept. of Astronomy, the Univ. of Michigan, 1085 S. University Ave., Ann Arbor, MI, 48109 \\
$^4$Dept. of Astronomy \& Astrophysics, Univ. of Chicago, 5640 S. Ellis Ave., Chicago, IL, 60637 \\
$^5$MPE, Postfach 1312, Giessenbachstr., D-85741 Garching, Germany \\  }

\pubyear{2014}
\volume{309}
\jname{Galaxies in 3D across the Universe}
\editors{B. L. Ziegler, F. Combes, H. Dannerbauer, M. Verdugo, eds.}

\begin{document}

\maketitle

\begin{abstract}
Star formation occurs on physical scales corresponding to individual star forming regions, typically of order $\sim$100 parsecs in size, but current observational facilities cannot resolve these scales within field galaxies beyond the local universe. However, the magnification from strong gravitational lensing allows us to measure the properties of these discrete star forming regions within galaxies in the distant universe. New results from multi-wavelength spectroscopic studies of a sample of extremely bright, highly magnified lensed galaxies are revealing the complexity of star formation on sub-galaxy scales during the era of peak star formation in the universe. We find a wide range of properties in the rest-frame UV spectra of individual galaxies, as well as in spectra that originate from different star forming regions within the same galaxy. Large variations in the strengths and velocity structure of Lyman-alpha and strong P Cygni lines such as C IV, and MgII provide new insights into the astrophysical relationships between extremely massive stars, the elemental abundances and physical properties of the nebular gas those stars ionize, and the galactic-scale outflows they power.
\end{abstract}

\firstsection
\section{Introduction}

The rest-frame ultraviolet spectra of galaxies contain a wealth of
information about the population of massive stars, the properties of 
the nebular gas those stars ionize, and the galactic-scale outflows they power.  
We can therefore use rest-UV spectra to constrain the famous ``galactic feedback''
that drives metals into the intergalactic medium and to better understand the role 
of this feedback in shutting down future star formation. A key science goal for 
20--30~m telescopes is to understand this feedback process, but until the next 
generation of telescopes are built, there are only two ways 
to obtain rest-UV diagnostics for typical star-forming galaxies at
redshifts above z $\sim$1.5. The first is to stack low-quality spectra of 
many galaxies (e.g., Shapley \etal\ 2003). This results in a good 
sampling of the average properties of star forming galaxies, but removes the 
possibility of understanding the variations in the observable properties and 
how changes in one or more observables affects some or all of the others.
The second way is to target galaxies that have been highly amplified by gravitational 
lensing. In previously published good signal-to-noise (S/N) rest-frame UV spectra for 
four such bright gravitationally lensed galaxies there are numerous strong features that trace 
the properties of massive stars and outflowing gas (Pettini \etal\ 2002, Finkelstein 
\etal\ 2009, Quider \etal\ 2009, 2010). All four of these galaxies have remarkably 
similar P~Cygni profiles, yet each has very different outflow 
properties -- we do not yet know what is a ``normal'' outflow at z $>$ 1.

At the heart of the problem is the fact that the fundamental physical scale of star formation 
and stellar outflows is not the scale of a single galaxy, but rather the scale of individual 
star forming regions within galaxies. However, strong gravitational lensing provides 
uniquely powerful views of distant galaxies, and in the most extreme 
high-magnification cases we are able to spatially resolute structures on $\lesssim$100 pc 
scales within galaxies at z $>$ 1. In this proceeding we present a few highlights of the 
first results from an ongoing effort to obtain good S/N rest-frame UV spectra 
of individual star forming regions within distant galaxies. Here we will focus on two 
source in particular -- RCS2 J0327-1326 and SGAS J1050+0017.

\begin{figure}
\centering
\includegraphics[width=0.98\columnwidth]{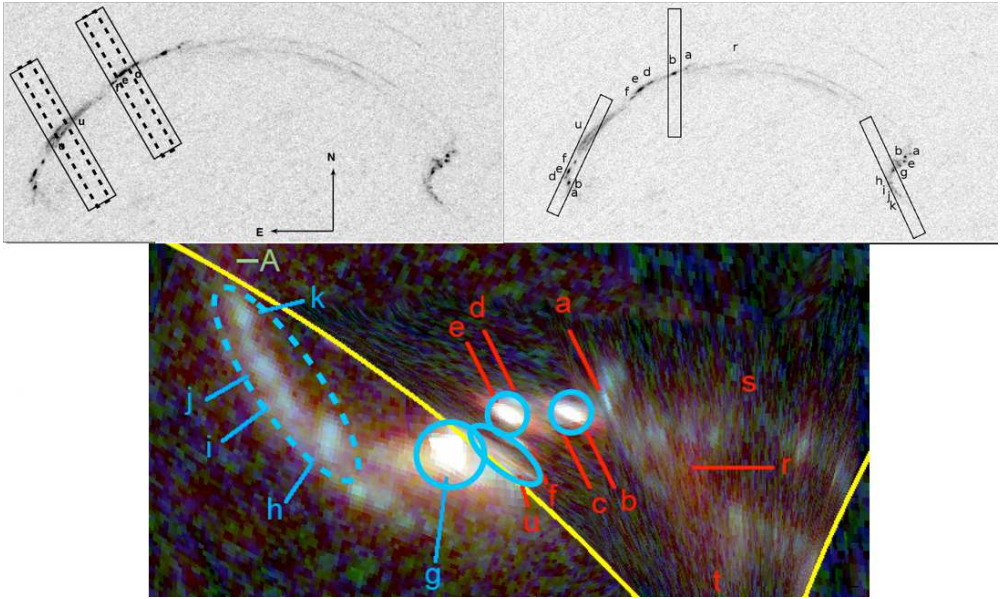} 
\caption{
Slit positions for MagE observations of RCS2 J0327-1326. The top grayscale images 
show individual emission regions/knots in the arc labeled with lowercase letters. Solid and 
dashed lines indicate the positions of the MagE slit along the arc. Slits indicated by 
solid and dashed lines on the top left depict the slit positions for observations taken 
at two different slit widths, where the wider-slit was used for observations during 
times of worse seeing. The bottom panel 
shows a reconstruction of the giant arc in the source plane from Sharon \etal\ (2012), 
with the same emission knots -- each with a physical size of $\lesssim$100 pc -- are 
labeled with lowercase letters. The solid cyan 
ellipsoids indicate the four knots for which we have good S/N  
spectra.}\label{fig:0327slits}
\end{figure}

\section{RCS2 J0327-1326}

RCS2 J0327-1326 is a spectacular strongly lensed galaxy at $z = 1.704$ that forms 
a giant arc extending $\sim$38'' along the sky (Wuyts \etal\ 2010, Sharon \etal\ 2012). 
This galaxy has been well-studied at NIR wavelengths, which samples the rest-frame optical 
(Rigby \etal\ 2011, Whitaker \etal\ 2014, Wuyts \etal\ 2014). We have also been 
conducting a follow-up campaign to acquire good S/N moderate resolution optical 
spectra of this arc with the MagE spectrograph on the Magellan-II (Clay) telescope. 
Due to the high magnification it is possible to place standard spectroscopic slits at different 
positions that map to distinct star forming regions with sizes $\lesssim$100 pc in the 
source galaxy (Figure~\ref{fig:0327slits}).

\begin{figure}
\centering
\includegraphics[width=0.496\columnwidth]{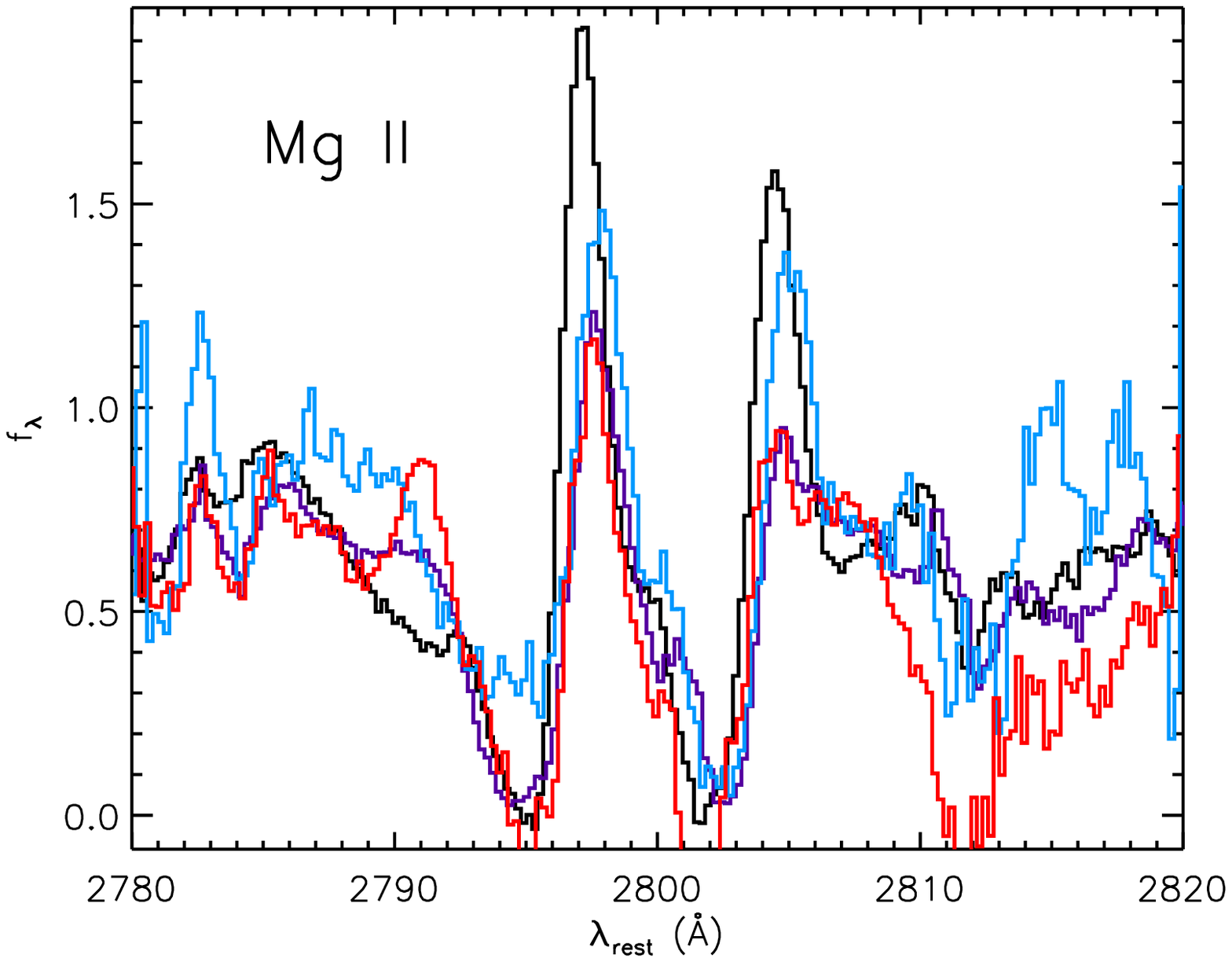} 
\includegraphics[width=0.496\columnwidth]{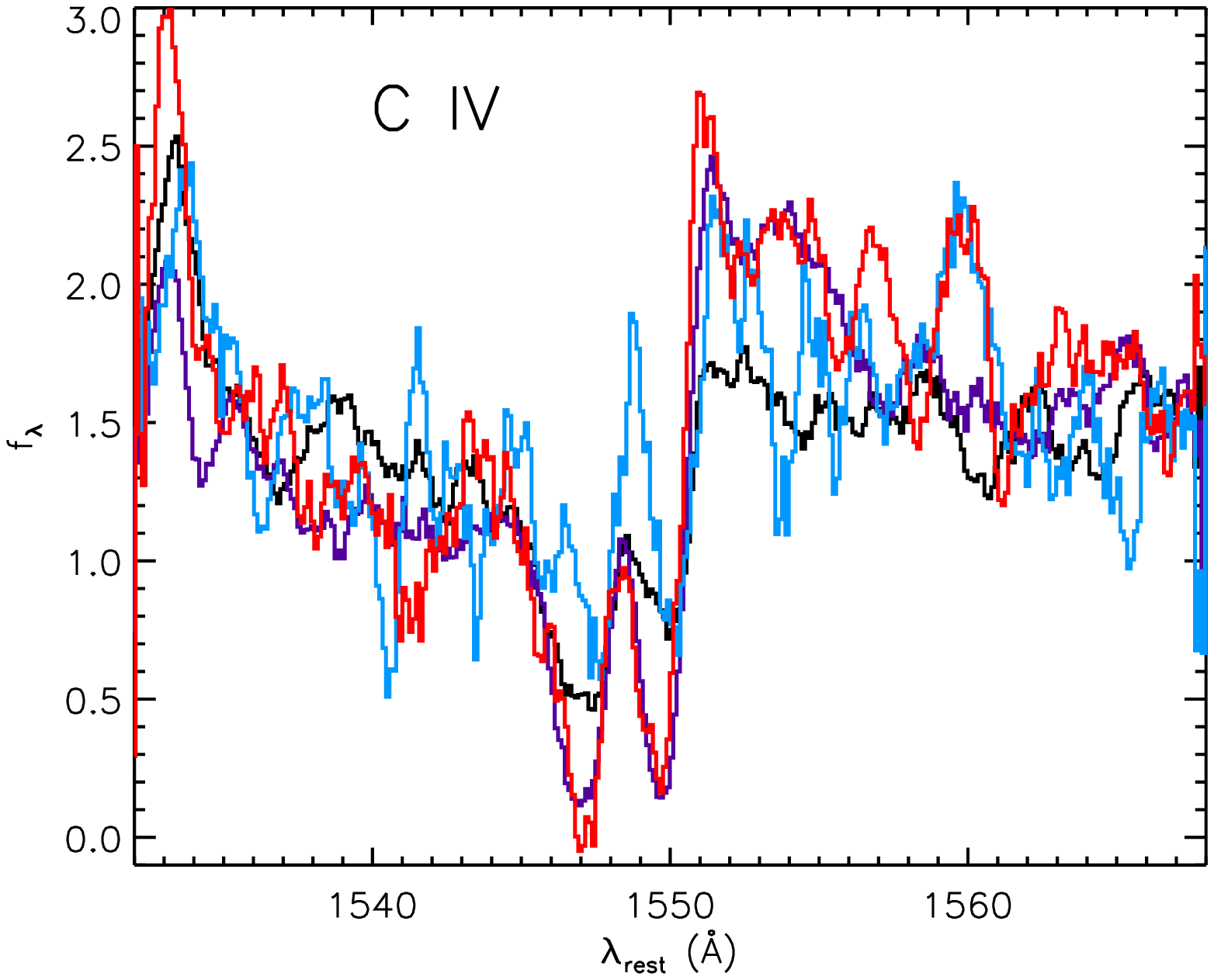} 
\caption{
The Mg {\small II} 2796,2803 (left) and C {\small IV} 1548,1551 (right) P Cygni lines 
as observed in four different $\lesssim$100 pc scale star forming regions within 
RCS2 J0327-1326. Both panels are consistent in their use of four 
different colors to indicate the four distinct star forming regions.
}\label{fig:0327_lines}
\end{figure}

Spectra of each of four star forming knots within RCS2 J0327-1326 
exhibit strong differences in the strength and structure of many prominent rest-UV 
features. In particular, the C {\small IV} and Mg {\small II} P Cygni lines, which result 
from outflowing gas, show large variations between the different knots 
(Figure~\ref{fig:0327_lines}). It is also notable that there is no positive correlation 
between the strength of the C {\small IV} and Mg {\small II} lines, indicating that 
these two lines are generated by different mechanisms and/or in different physical 
locations. These data also informed the observed lack of correlation between 
Ly-$\alpha$ and Mg {\small II} P Cygni emission in a sample of strongly lensed galaxies 
(Rigby \etal\ 2014), which further supports a physical picture in which Mg {\small II} emission 
traces stellar wind driven outflows, possibly providing a diagnostic measure of the 
radiative transfer mechanisms in those outflows.

\section{SGAS J1050+0017}

\begin{figure}
\centering
\includegraphics[width=0.496\columnwidth]{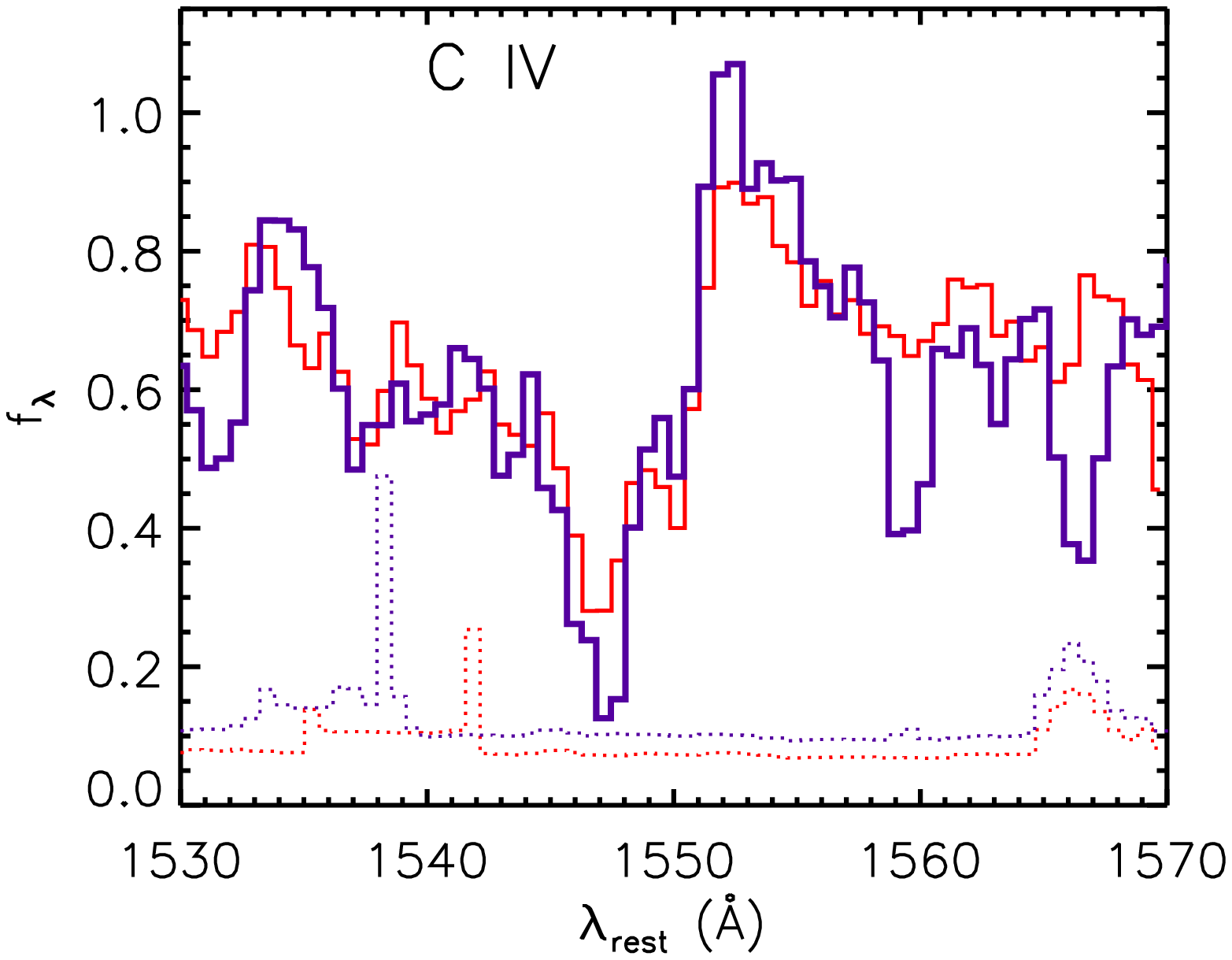}
\includegraphics[width=0.496\columnwidth]{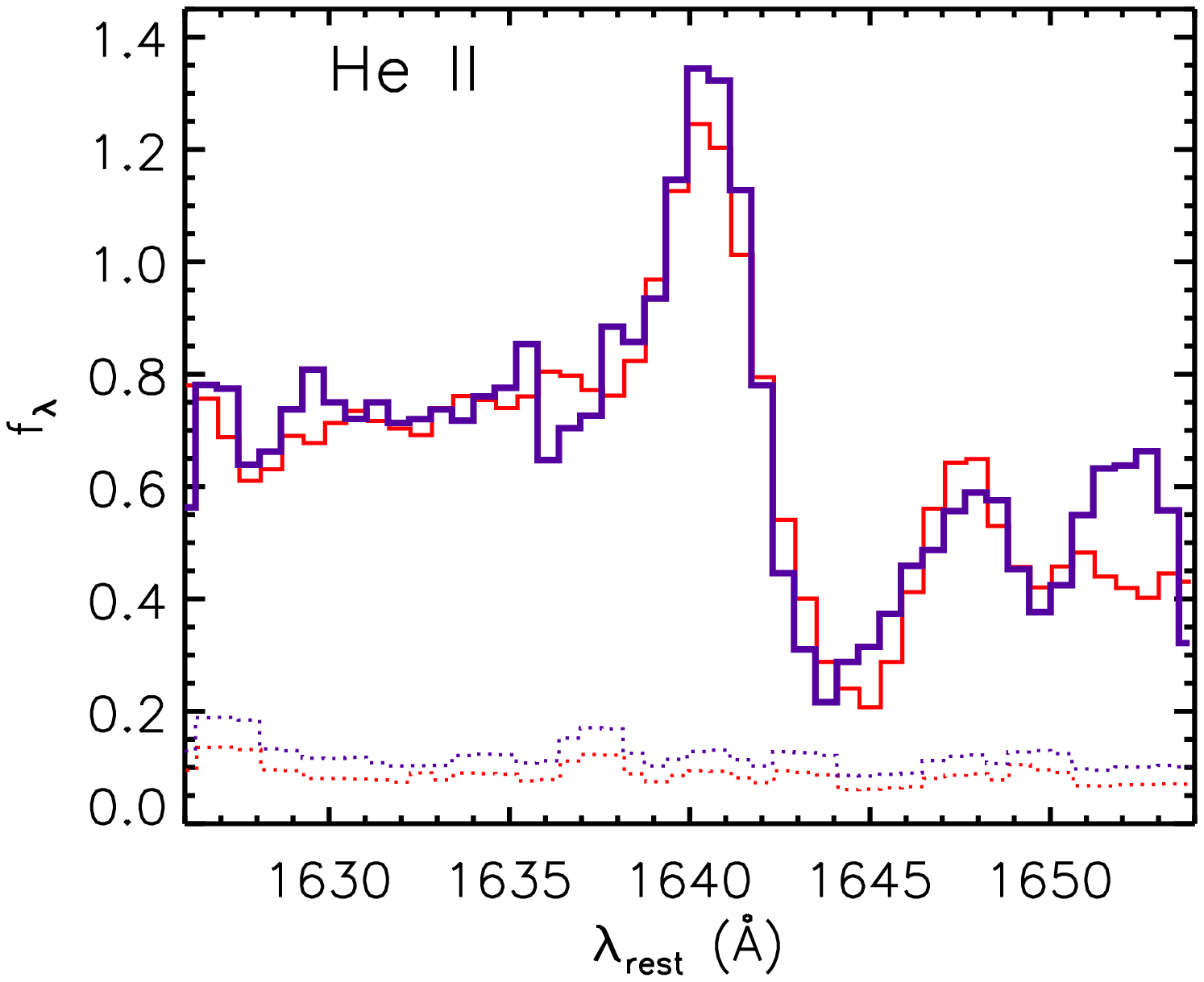}
\caption{
The C {\small IV} 1548,1551 (left) P~Cygni lines and He {\small II} 1640 (right) emission 
line as seen in distinct star forming regions (indicated by different colors) within 
SGAS J1050+0017.
}\label{fig:s1050}
\end{figure}

This bright lensed galaxy at $z = 3.625$ was published by Bayliss \etal\ (2014), in which 
the authors analyzed optical through NIR (rest-frame UV through optical) spectra of the 
source, as well as imaging data spanning 0.4 through 4.5 microns. This multi-wavelength 
analysis had difficultly explaining the properties of the integrated spectra of the giant arc, 
including very strong and narrow P Cygni features, as well as differences 
between the ionization parameter measurements. Here we show a new analysis of the 
Gemini/GMOS-North spectra of this arc in which we isolate and individually extract the 
emission from two distinct $\sim$100-200 pc star forming knots along the arc (see Figures 
1 \& 2 in Bayliss \etal\ 2014). The spectra clear exhibit differences between the two 
distinct knots, with the second knot, plotted in red, having a much weaker C {\small IV} 
feature, as well as weaker He {\small II} emission (Figure~\ref{fig:s1050}). These spectra 
provide direct evidence that the conditions of the ionizing radiation field and the population 
of massive stars vary significantly across different regions at $z = 3.625$.

\section{Conclusions}


Looking at individual star forming regions within the most highly magnified giant 
arcs we see significant variations along different lines of sight, including variable 
strength and profile shape of structure of P Cygni wind lines, with no evidence for 
correlation between C {\small IV} and Mg {\small II} P Cygni features, and 
indications of significant differences in the ionizing radiation field. These results 
reinforce the argument that the astrophysics of the interstellar medium (ISM) 
in distant star forming galaxies is a complex business. 

Bright lensed galaxies extending out to z $\sim$5 have recently become common place 
(e.g., Bayliss \etal\ 2010, Koester \etal\ 2010, Bayliss \etal\ 2011b, Gladders \etal\ in prep),
and these lensed galaxies typically reside at z $\sim$ 2, and therefore sample the 
epoch of peak star formation in the universe (Bayliss \etal\ 2011a, Bayliss 2012). 
The highest magnification systems drawn from these large samples provide a new 
opportunity to leave single-object analysis behind and study the UV spectra of 
individual star forming regions within many z $>$ 1 galaxies. Such observations 
would allow us to characterize the relationship between local signatures of the winds 
of massive stars and their interaction with their surrounding ISM. These data will, in 
turn, reveal important new information about the astrophysics of the radiative transfer 
in the ISM in regions of prolific star formation. 

\section*{Acknowledgements}
\noindent
This work is based on observations from the Magellan-II (Clay) and the Gemini-North 
Telescopes. Supported was provided by the National Science Foundation through Grant 
AST-1009012, by NASA through grant HST-GO-13003.01.

\end{document}